\documentclass[10pt,conference]{IEEEtran}
\usepackage{epsfig,graphicx,subfigure,psfrag,amsmath,cases,bm}
\usepackage{latexsym,amssymb,algorithm,mathtools}
\usepackage{algorithmic}
\usepackage{color}
\usepackage{url}
\usepackage{scrtime}
\usepackage{stfloats}
\usepackage{tablefootnote}
\usepackage{cite}
\usepackage{amsfonts,mathabx}
\usepackage{textcomp}
\usepackage{amsthm}
\usepackage{xcolor,capt-of}
\usepackage{multirow}
\usepackage[left=0.58in, right=0.58in, bottom=0.98in, top=0.7in]{geometry}
\makeatletter
\newcommand\scalemath[2]{\scalebox{#1}{\mbox{\ensuremath{\displaystyle #2}}}}
\newcommand*{\rom}[1]{\expandafter\@slowromancap\romannumeral #1@}
\makeatother

\DeclareMathOperator{\mino}{minimize}

\newtheorem{lemma}{Lemma}

\allowdisplaybreaks

\def\BibTeX{{\rm B\kern-.05em{\sc i\kern-.025em b}\kern-.08em
T\kern-.1667em\lower.7ex\hbox{E}\kern-.125emX}}
\title{Advanced ISAC Design: Movable Antennas and Accounting for Dynamic RCS}
\author{\IEEEauthorblockN{Ata Khalili and Robert Schober\\
Friedrich-Alexander-University Erlangen-Nurnberg, Germany}\vspace{-10mm}}

\begin{document}
\maketitle
\begin{abstract}
We investigate resource allocation in integrated sensing and communication (ISAC) systems exploiting movable antennas (MAs) to enhance system performance. Unlike the existing ISAC literature, we account for dynamic radar cross-section (RCS) variations. Chance constraints are introduced and integrated into the sensing quality of service (QoS) framework to precisely control the impact of the resulting RCS uncertainties. Taking into account the dynamic nature of the RCS, we jointly optimize the MA positions and the communication and sensing beam design for minimization of the total transmit power at the base station (BS) while ensuring the individual communication and sensing task QoS requirements. To tackle the resulting non-convex mixed integer non-linear program (MINLP), we develop an iterative algorithm to obtain a high quality suboptimal solution. Our numerical results reveal that the proposed MA-enhanced ISAC system cannot only significantly reduce the BS transmit power compared to systems relying on fixed antenna positions and antenna selection but also demonstrates remarkable robustness to RCS fluctuations, underscoring the multifaceted benefits of exploiting MAs in ISAC systems.
\end{abstract}
\section{Introduction}
In the realm of sixth-generation (6G) wireless networks, the significance of multiple-input multiple-output (MIMO) systems for enhanced data rates, security, and integrated sensing and communication (ISAC) is undeniable\cite{mietzner2009multiple,Globecom2023}. However, conventional MIMO systems, which are based on multiple radio frequency (RF) chains, entail high hardware costs and complexity\cite{mietzner2009multiple}. To address these issues, antenna selection (AS) has emerged as a practical solution, aiming to reduce the number of RF chains by selecting a subset of antennas with favorable channel characteristics\cite{sanayei2004antenna}. Traditional MIMO systems, whether they employ AS or not, are constrained by the fixed positions of their antennas. This fixed positioning cannot fully leverage the channel variations across the spatially continuous transmitter area. Utilizing these variations could potentially enhance system performance by facilitating diversity and multiplexing gains. Holographic MIMO enables the exploitation of the channel variations across the transmitter area by employing densely packed, electronically controlled passive elements. However, the practical implementation of holographic MIMO is challenging because of the complexity and cost associated with managing a large array of narrowly-spaced antenna elements, which complicates channel estimation and data processing\cite{huang2020holographic}.

Inspired by the spatial degrees of freedom (DoFs) offered by holographic MIMO surfaces, the novel MIMO concept of movable antennas (MAs) and fluid antennas has emerged as a bridge between holographic MIMO and traditional MIMO \cite{zhu2022modeling,fluidantenna}. In MA-enabled systems, each antenna element is connected to an RF chain via a flexible cable, allowing physical adjustments within a designated spatial region using electromechanical devices like stepper motors \cite{ma2022mimo}. This mobility allows the optimization of the antenna positions to achieve favorable spatial antenna correlations, significantly enhancing the MIMO system's capacity. Movable antennas (and similarly also fluid antennas) provide greater spatial flexibility compared to fixed-position antennas, resulting in enhanced system performance while requiring fewer antenna elements to exploit the available DoFs\cite{zhu2022modeling, ma2022mimo, zhu2023movable}.

To tap into the full potential of MA-enabled systems, initial works have explored joint beamforming and antenna positioning. In \cite{ma2022mimo}, an alternating optimization (AO) algorithm was proposed for MA-enabled MIMO systems, where the base station (BS) and multiple users were equipped with MAs. Similarly the authors of \cite{zhu2023movable} considered a multiuser MA-enabled uplink communication system with a fixed antenna array at the BS. Employing zero-forcing (ZF) or minimum mean square error (MMSE) combining, the MA positions were adjusted using a gradient descent (GD) method. However, both works optimistically assumed continuous adjustment of the MA positions within a given region, which may not be practical. Prototype designs in \cite{zhuravlev2015experimental} and \cite{basbug2017design} employ discrete motion control of electromechanical devices with finite precision, leading to a quantized transmitter area and finite spatial resolution.

In recent years, ISAC has attracted significant attention as a promising technology to increase spectrum efficiency and enable the sharing of physical infrastructure for information transmission and environmental data collection\cite{ISAC6G}. In this regard, the authors of \cite{mu-mimo-jsc,jsc-mimo-radar} studied transmit beamforming for MIMO ISAC systems, where a least-squares problem was formulated to obtain the ideal beampattern for sensing, while guaranteeing the quality of service (QoS) of the communication users. However, the achieved signal-to-noise ratio (SNR) of the received radar echoes was not considered, which may lead to unreliable sensing. Furthermore, although the dynamic nature of the radar cross-section (RCS) is well documented in the radar literature, and captured by the Swerling models \cite{Radar}, it has been largely overlooked in the ISAC literature. In this paper, we bridge this gap and take into account the uncertainty caused by the dynamic nature of the RCS by modeling its impact on the sensing SNR.

Furthermore, we reveal that MAs offer a unique advantage for ISAC systems. The dynamic reconfigurability and sub-wavelength positioning of the antenna elements enhances both  data transmission and sensing accuracy by enabling precise beamforming and optimal beamwidth design, avoiding undesirable side lobes, and improving interference management. Crucially, the flexible sub-wavelength positioning of the MAs is vital for managing the dynamic RCS. 
This paper pioneers the exploitation of MAs with a spatially discrete transmitter area and the consideration of the uncertainty introduced by the dynamic RCS for enhancing ISAC system performance. The main contributions of this paper can be summarized as follows:
\begin{itemize}
    \item We investigate the benefits of MAs for ISAC systems, where we account for the discrete nature of the possible MA positions.
    \item  We introduce a new sensing performance metric leveraging chance constraint to address fluctuations in the SNR caused by the dynamic RCS of the sensing targets. 
    \item We jointly optimize the MA positions and the downlink information and sensing beamformers for minimization of the transmit power of the BS, which results in a challenging non-convex mixed integer non-linear program (MINLP).
    \item Our simulations demonstrate that strategic MA positioning can significantly reduce power consumption in ISAC systems compared to baseline methods. Furthermore, expanding the transmitter area enhances performance, albeit at the expense of increased computational complexity.
\end{itemize}
\textit{Notation:} 
In this paper, matrices and vectors are denoted by
boldface capital letters $\mathbf{A}$ and lower case letters $\mathbf{a}$, respectively.~$\mathbf{A}^T$, $\mathbf{A}^{*}$, $\mathbf{A}^H$, $\text{Rank}(\mathbf{A})$, and $\text{Tr}(\mathbf{A})$ are the transpose,~conjugate, Hermitian, rank, and trace of matrix $\mathbf{A}$, respectively. 
$\mathbf{A}\succeq\mathbf{0}$ denotes a positive semidefinite matrix. $\mathbf{I}_N$ is the $N$-by-$N$ identity matrix. $\mathbb{R}^{N\times M}$ and $\mathbb{C}^{N\times M}$ represent the space of $N\times M$ real-valued and complex-valued matrices, respectively. $|\cdot|$ and $||\cdot||_2$ stand for the absolute value of a complex scalar and the $l_2$-norm of a vector, respectively. $\mathbf{0}_{L}$ and $\mathbf{1}_L$ represent the all-zeros and all-ones vectors of length $L$, respectively. $\mathbb{E}[\cdot]$ refers to statistical expectation.
\section{System Model}
\begin{figure}
    \centering
    \includegraphics[width=3.4in]{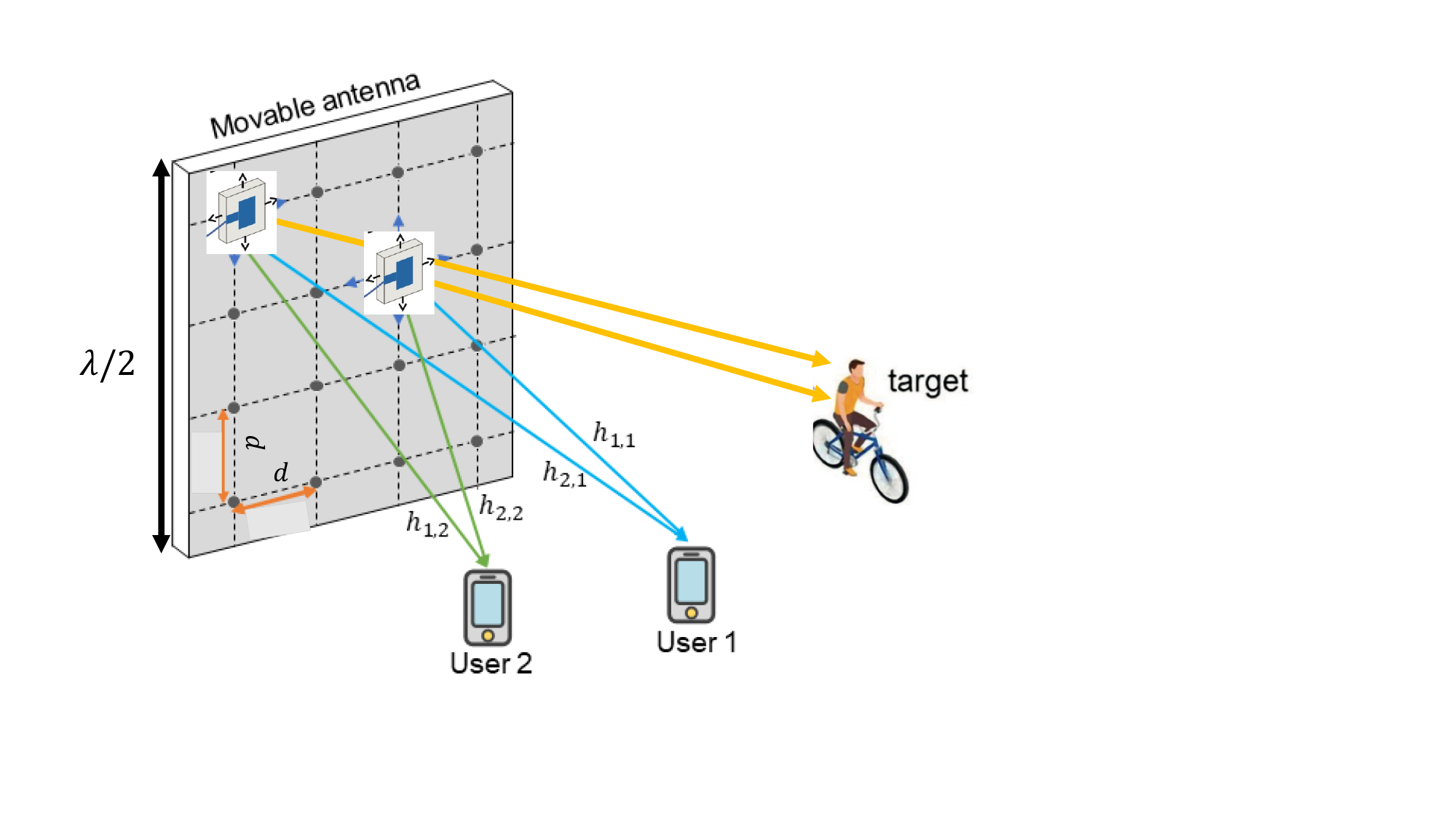}
    \vspace{-10mm}
    \caption{$N=2$ movable antenna elements with $M=16$ possible discrete positions transmit to $K=2$ users and $E=1$ sensing target.}
    \label{fig:MA_system_model}
    \vspace{-2mm}
\end{figure}
We consider a dual function radar and communication (DFRC)-BS that is equipped with $N$ MA elements, capable of sub-wavelength positioning to enhance spatial resolution. The DFRC-BS  serves $K$ single-antenna communication users and concurrently senses $E$ potential targets with high precision. The MA elements can be simultaneously adjusted within a designated two-dimensional transmitter area. We assume that perfect channel state information (CSI) is available at the DFRC-BS in order to reveal the maximum performance of MA-assisted ISAC systems\cite{ma2022mimo, zhu2023movable}. 
\subsection{Transmitter Model}
The transmitter area of the MA-enabled communication and sensing system is quantized\cite{basbug2017design}.
We collect the $M$ possible discrete positions of the MAs in set $\mathcal{P}=\{\mathbf{p}_1,\cdots, \mathbf{p}_{M}\}$, where the distance between the neighboring positions is equal to $d$ in horizontal and vertical direction, as shown in Fig. \ref{fig:MA_system_model}. Here, $\mathbf{p}_m=[x_m,y_m]$ represents the $m$-th candidate position with horizontal coordinate $x_m$ and vertical coordinate $y_m$. 
In other words, the feasible set of the position of the $n$-th MA element, $\mathbf{t}_{n}$, is given by $\mathcal{P}$, i.e., $\mathbf{t}_n\in\mathcal{P}$. For notational simplicity, we define sets $\mathcal{K}=\{1,\cdots,K\}$, $\mathcal{N}=\{1,\cdots,N\}$, and $\mathcal{M}=\{1,\cdots,M\}$ to collect the indices of the users, MA elements, and candidate positions of the MA elements, respectively. Furthermore, we define the binary position selection vector of the $n$-th MA element as $\mathbf{b}_n=\big[b_n[1],\cdots,b_n[M]\big]^T,\hspace*{1mm}\forall m$, where $b_n[m]\in\left\{0,\hspace*{1mm}1\right\}$ and $\sum_{m=1}^{M}b_n[m]=1$, $\forall n$. Note that $b_n[m]=1$ if and only if the $m$-th discrete position in $\mathcal{P}$ is selected for the $n$-th MA element \cite{Yifi}. 
As antenna elements cannot be infinitely small, two MA elements cannot be positioned arbitrarily close to each other. Therefore, the center-to-center distance between any two MA elements must exceed a certain minimum distance, $D_{\min}$. We define a distance matrix $\mathbf{D}\in\mathbb{C}^{M\times M}$, whose entry $D_{m,m'}$ represents the distance between the $m$-th and $m'$-th candidate positions in $\mathcal{P}$. Thus, the minimum distance constraint between any pair of MA elements can be expressed as follows:
\begin{equation}
\mathbf{b}_n^T\mathbf{D}\mathbf{b}_{n'}\geq D_{\mathrm{min}},\ n\neq n',\ \forall n, n'\in\mathcal{N}.
\end{equation}
\subsection{Signal Model}
     The DFRC-BS transmits simultaneously information symbols $c_{k}\sim \mathcal{CN}(0,1)$, $k \in \{1,...,K\}$, to $K$ communication users. In addition, for sensing, dedicated radar signal $\mathbf{s}_{0} \in \mathcal{C}^{N \times 1}$ with covariance matrix $\mathbf{R} = \mathbb{E}[\mathbf{s}_0\mathbf{s}_0^H] \succeq \mathbf{0}$ is also concurrently transmitted. Here,  communication and radar signals are assumed to be statistically independent such that $ \mathbb{E}[c_k^{*}\mathbf{s}_0]=\mathbf{0}$. The baseband transmit signal of the DFRC-BS can be expressed as follows
     \vspace{-2mm}
\begin{equation}\label{trans}
\mathbf{x}= \sum_{k=1}^K \mathbf{w}_k c_k+\mathbf{s}_{0},
\vspace{-3mm}
\end{equation}
where $\mathbf{w}_k \in \mathbb{C}^{N \times 1}$ denotes the transmit beamforming vector for user $k$. To facilitate the sensing of multiple targets, we employ a multi-beam signal. Thus, $\mathbf{s}_{0}$ is given by
$\mathbf{s}_0 = \sum_{n=1}^{N} \mathbf{v}_{n} s_{n}$,
where $s_{n}$ are independent and identically distributed pseudo-random symbols, having zero mean and unit variance, and $\mathbf{v}_n$ is the $n$-th sensing beamforming vector. Accordingly, the sensing beamformers $\{\mathbf{v}_{n}\}_{n=1}^{N}$ can be obtained from $\mathbf{R}$ via eigenvalue decomposition (EVD).  Based on \eqref{trans}, the covariance matrix of the transmit signal is given by $\mathbf{R}_{\mathbf{x}}=\mathbb{E}[\mathbf{x}\mathbf{x}^{H}]=\sum_{k=1}^K \mathbf{w}_k \mathbf{w}^{H}_k+\mathbf{R}$. 
\subsection{Communication Channel and Metric}
In the considered MA-enabled MIMO system, the physical channel can be reconfigured by adjusting the positions of the MA elements. The channel vector between the $n$-th MA element and the $K$ users is denoted by $\mathbf{h}_n(\mathbf{t}_n)=[h_{n,1}(\mathbf{t}_n),\cdots,h_{n,K}(\mathbf{t}_n)]^T$ and depends on the position of the $n$-th MA element, $\mathbf{t}_n$, where $h_{n,k}(\mathbf{t}_n)\in\mathbb{C}$ denotes the channel coefficient between the $n$-th MA element and the $k$-th user. The channel between the $m$-th candidate position of the $n$-th MA element, $\mathbf{p}_{m}$, and the $k$-th user is given by $h_{n,k}(\mathbf{p}_m)$, which is modelled as 
$h_{n,k}(\mathbf{p}_m)=\mathbf{1}_{L_p}^T\bm{\Sigma}_k\mathbf{g}_k(\mathbf{p}_m)$,
where $\mathbf{1}_{L_p}$ denotes the uniform field response vector (FRV) at the $k$-th user, which has a single, non-adjustable antenna \cite{ma2022mimo,zhu2022modeling}. The diagonal matrix $\bm{\Sigma}_k=\mathrm{diag}{[\sigma_{1,k},\cdots,\sigma_{L_p,k}]^T}$ contains on its main diagonal the path weights of the $L_p$ paths that extend from the transmitter's location to the $k$-th user. The path weights $\sigma_{l_p,k}$, $l_p\in \{1,...,L_{p}\}$, are independently distributed, and follow complex Gaussian distribution $\mathcal{CN}(0,L_0D_{l_{{p},k}}^{-\alpha})$. Here, $L_0$ represents the reference large-scale fading at a distance of $d_0=1$ m, $D_{l_{{p},k}}$ is the distance between the BS and the $k$-th user via the $l_p$-th scatterer, and $\alpha$ denotes the path loss exponent. Furthermore, $\mathbf{g}_k(\mathbf{p}_m)$ denotes the transmit FRV linking the $k$-th user to the $m$-th MA position, $\mathbf{p}_m$, and is given by 
$\mathbf{g}_k(\mathbf{p}_m)=\left[e^{j\rho_{k,1}(\mathbf{p}_m)},\cdots,e^{j\rho_{k,L_p}(\mathbf{p}_m)}\right]^T$,
where $\rho_{k,l_p}(\mathbf{p}_m)=\frac{2\pi}{\lambda}\Big((x_m-x_1)\cos\theta_{k,l_p} \sin\phi_{k,l_p}+$
$(y_m-y_1)\sin\theta_{k,l_p}\Big)$ represents the phase difference between $\mathbf{p}_m$ and the first position $\mathbf{p}_1$ for the $l_p$-th channel path, and $\lambda$ is the carrier wavelength \cite{ma2022mimo,zhu2022modeling}. For the $k$-th user, $\theta_{k,l_p}$ and $\phi_{k,l_p}$ represent the elevation and azimuth angles of departure (AoD), respectively, for the $l_p$-th channel path. The distribution of the angles is assumed to be given by $f_{\mathrm{AoD}}(\theta_{k,l_p}, \phi_{k,l_p})=\frac{\cos\theta_{k,l_p}}{2\pi}$, where both $\theta_{k,l_p}$ and $\phi_{k,l_p}$ are in the range of $[- \pi/2$, $\pi/2]$ \cite{zhu2023movable}.  Next, we define matrix $\hat{\mathbf{H}}_{n}=[\mathbf{h}_{n}(\mathbf{p}_1),\cdots,\mathbf{h}_{n}(\mathbf{p}_M)]\in\mathbb{C}^{K\times M}$ collecting the channel vectors from the $n$-th MA element to all $K$ users for all $M$ feasible discrete MA locations. Then, $\mathbf{h}_{n}(\mathbf{t}_n)$ can be expressed as $\mathbf{h}_{n}(\mathbf{t}_n)=\hat{\mathbf{H}}_{n}\mathbf{b}_n$.
For the considered MA-enabled multiuser MISO system, the channel matrix between the BS and the $K$ users, $\mathbf{H}=[\mathbf{h}_{1}(\mathbf{t}_1),\cdots,\mathbf{h}_{N}(\mathbf{t}_N)]\in\mathbb{C}^{K\times N}$, is then given by
  $ \mathbf{H}=\hat{\mathbf{H}}\mathbf{B}$,
where matrices $\hat{\mathbf{H}}\in \mathbb{C}^{K\times MN}$ and $\mathbf{B}\in \mathbb{C}^{MN\times N}$ are defined as follows, respectively,
\begin{eqnarray}
\hat{\mathbf{H}}&\hspace*{-2mm}=\hspace*{-2mm}&
[\hat{\mathbf{H}}_{1},\cdots,\hat{\mathbf{H}}_{N}],\\
\mathbf{B}&\hspace*{-2mm}=\hspace*{-2mm}&
  \begin{bmatrix}
    \mathbf{b}_1 & \mathbf{0}_{M} & \mathbf{0}_{M} & \cdots & \mathbf{0}_{M}\\
    \mathbf{0}_{M} & \mathbf{b}_2 & \mathbf{0}_{M} &\cdots & \mathbf{0}_{M}\\
    \ldots & \ldots & \ldots & \ldots & \ldots\\
    \mathbf{0}_{M} & \mathbf{0}_{M} & \mathbf{0}_{M} & \hspace*{1mm}\cdots & \mathbf{b}_N
  \end{bmatrix}.
\end{eqnarray}
Next, we define $\hat{\mathbf{h}}_k\in\mathbb{C}^{1\times MN}$ as the $k$-th row of $\hat{\mathbf{H}}$. Then, the received signal of the $k$-th user is given by
$y_{k}=\hat{\mathbf{h}}_k\mathbf{B}\sum_{l\in \mathcal{K}}\mathbf{w}_{l}{c}_{l}+\hat{\mathbf{h}}_k\mathbf{B}\sum_{n\in \mathcal{N}}\mathbf{v}_{n}s_{n}+n_{k}$,
where $n_k\in\mathbb{C}$ denotes the additive white Gaussian noise (AWGN) at the $k$-th user with zero mean and variance $\sigma_k^2$. Thus, the signal-to-interference-plus-noise ratio (SINR) of the $k$-th user is given by 
\begin{align}
 \gamma_k=\frac{|\hat{\mathbf{h}}_{k}\mathbf{B}\mathbf{w}_{k}|^{2}}{\sum_{i\in\mathcal{K}\setminus\{k\}}|\hat{\mathbf{h}}_{k}\mathbf{B}\mathbf{w}_{i}|^2+\hat{\mathbf{h}}_{k}\mathbf{B}\mathbf{R}\mathbf{B}^{T}\hat{\mathbf{h}}^{H}_{k}+\sigma_{k}^2}.   
\end{align}
\subsection{Sensing Channel and Metrics}
For target $e$, the FRV of the $n$-th MA element for all $M$ feasible discrete MA positions is given by
$\mathbf{a}_{n}(\theta_{e},\phi_{e})\hspace{-0.25mm}=\big[e^{j \rho_{e}(\mathbf{p}_{1}) },\cdots,e^{j\rho_{e}(\mathbf{p}_{M}) } \big]^T$,
where $\rho_{e}(\mathbf{p}_{m})=\frac{2\pi}{\lambda}\Big((x_{m}-x_{1})\cos\theta_{e}\sin\phi_{e}+(y_{m}-y_{1})\sin\theta_{e}\Big)$, and $\theta_{e}$ and $\phi_{e}$ are the elevation and azimuth AoDs corresponding to target $e$. Next, we concatenate the individual FRVs of all MA antenna elements, which leads to 
$\mathbf{\hat{a}}(\theta_{e}, \phi_{e}) =\big[ 
\mathbf{a}^{T}_{1}( \theta_{e}, \phi_{e}), 
\cdots, \
\mathbf{a}^{T}_{N}(\theta_{e}, \phi_{e})\big]^{T}$.
The steering vector $\mathbf{a}(\theta_{e}, \phi_{e})=\big[e^{j \rho_{e}(\mathbf{t}_{1}) },\cdots,e^{j\rho_{e}(\mathbf{t}_{N}) } \big]^T$ of the $N$ MA elements can thus be expressed as $\mathbf{a}(\theta_{e}, \phi_{e})=\mathbf{B}^{T}\mathbf{\hat{a}}(\theta_{e}, \phi_{e})$.
\subsubsection{Beam Pattern Matching Design}
To ensure high-quality sensing, the desired target locations have to be illuminated by an energy-focusing beam with low side lobe leakage such that the desired echoes can be easily distinguished from clutter. To this end, we discretize the elevation angle domain $[-\frac{\pi}{2}, \frac{\pi}{2}]$ to $L$ directions and the azimuth domain $[-\frac{\pi}{2}, \frac{\pi}{2}]$ to $Q$ directions \\ \\  and specify the ideal beam pattern $\{\mathcal{D}{(\theta_l,\phi_{q})}\}_{q=1,l=1}^{Q,L}$,  where  $\mathcal{D}(\theta_l,\phi_{q})$ is given by
\vspace{-2mm}
\[
\mathcal{D}{(\theta_l,\phi_q)}=
\begin{cases}\label{desired_a}
1, &\theta_{e}-\Delta\leq \theta_l\leq\theta_{e}+\Delta \\ &\text{and} ~~\phi_{e}-\delta\leq \phi_{q} \leq\phi_{e}+\delta, e=1,\ldots,E \:\:\\
  0, & \text{otherwise},
\end{cases}
\]
where $2\Delta$ and $2\delta$ are the beamwidths for each target with respect to the elevation and azimuth angles, respectively \cite{Probing}. Consequently, to quantify the beam pattern matching accuracy, we adopt the mean square error (MSE) between the ideal beam pattern and the actual beam pattern as a performance metric for sensing \cite{Probing}, which is given by
\begin{align}\label{Rd}
\scalemath{0.95}{\frac{1}{Q} \frac{1}{L} \sum_{q=1}^Q\sum_{l=1}^L \bigg|\rho_{0}\mathcal{D}(\theta_l,\phi_{q}) -\mathbf{\hat{a}}^H(\theta_l,\phi_{q}) \mathbf{B}\mathbf{R_{x}}\mathbf{B}^{T}\mathbf{\hat{a}}(\theta_l,\phi_{q})\bigg|^2},
 \end{align}
 where $\rho_{0}$ is a scaling factor.
 
\subsubsection{Received Echo Signal}The
communication waveform is completely known at the DFRC-BS and thus its reflected signal can
also be exploited for target detection. Thus, the communication signal is not treated as interference at the radar receiver. Under the assumption that the transmit waveform is narrow-band and the sensing channel is line of sight (LoS) \cite{jsc-mimo-radar}, the received echo signal at the DFRC-BS is given by
$\mathbf{r}_{e}= {\mathbf{H}_{e}
 \mathbf{x}}+ \mathbf{z}$,
   where  $\mathbf{H}_{e}\hspace{-1mm}=\frac{\epsilon_{e}L_0}{2\Psi_{e}}  \mathbf{{a}}(\theta_{e},\phi_{e})\mathbf{{a}}^{H}(\theta_{e},\phi_{e})$ is the round-trip channel matrix of the $e$-th target and $\mathbf{z}\sim\mathcal{C}\mathcal{N}(\mathbf{0},\sigma^{2}_{e}\mathbf{I}_{N})$ is the received AWGN at the BS.
Here, $\Psi_{\mathrm{e}}$ denotes the distance between the DFRC-BS and target $e$, $\epsilon_{e}= \sqrt{\frac{\Omega_{e}}{4\pi \Psi^2_e}}$ is the reflection coefficient of target $e$, and $\Omega_{e}$ is the radar cross-section of target $e$. In contrast to the existing ISAC literature, where the RCS is assumed to be constant and known, we account for the dynamic nature of the RCS, characterized by the Swerling model in the radar literature\cite{Radar}. In particular, $\Omega_{e}$ is modeled as exponentially distributed with probability density function (PDF)
$p(\Omega_{e})=\frac{1}{\Omega_{\text{av}}}\text{exp}(\frac{-\Omega_{e}}{\Omega_{\text{av}}})$,
where $\Omega_{\text{av}}$ is the average RCS \cite{Radar}. To achieve satisfactory sensing performance, we require the sensing SNR of target $e$ to be higher than a pre-defined minimum threshold. This requirement is mathematically expressed as
${\Gamma_{e}\triangleq\frac{\Omega_{e}L_{0}^{2}\mathbf{\hat{a}}^H(\theta_{e},\phi_{e})\mathbf{B}\mathbf{R_{x}}\mathbf{B}^{T}\mathbf{\hat{a}}(\theta_{e},\phi_{e})}{16\pi \Psi^4_{e}\sigma^{2}_{e}}}>\Gamma^{\text{th}}_{e}$,
where $\Gamma^{\text{th}}_{e}$ is the minimum SNR required at the BS for sensing target $e$.
Given the dynamic nature of the RCS fluctuations, uncertainties for system design arise. In response to these uncertainties, we adopt a chance constraint to be able to provide performance guarantees. To this end, we impose the following probabilistic constraint to maintain the QoS for the sensing target
\vspace{-2mm}
\begin{align}\label{Chance}
\text{Pr}\bigg \{\Gamma_{e}\leq\Gamma^{\text{th}}_{e}\bigg \}\leq \nu,
\end{align} where $\nu$, $0<\nu<1$, denotes the maximum tolerable probability of failure. By enforcing \eqref{Chance}, we ensure that despite the uncertainty imposed by the dynamic RCS, the desired sensing SNR is achieved at least with probability $1-\nu$. 
\section{Problem Formulation and Proposed Solution}
\subsection{Problem Formulation}
In this paper, we aim to minimize the power consumption of the considered system while guaranteeing the QoS for each communication user and sensing target. \\ \\ The resulting resource allocation problem is formulated as follows:
\begin{align}
\label{Ori_Problem}
  &\hspace*{-4mm}\mathcal{P}_{0}:\underset{\mathbf{B},\mathbf{w}_{k},\mathbf{R},\rho_{0}}{\mino}\hspace*{2mm}\mathcal{F}\triangleq\sum_{k\in\mathcal{K}}\left\|\mathbf{w}_k\right\|_2^2\notag+\text{Tr}(\mathbf{R})\notag\\
    \mbox{s.t.}~~~&\mbox{C1:}~~\gamma_{k}\geq \gamma^{\text{th}}_{k},\hspace*{1mm}\forall k\in \mathcal{K},\notag\nonumber\\
&\mbox{C2:}\sum_{m=1}^Mb_n[m]=1,~
    \mbox{C3:}~b_n[m]\in \{0,1\},\scalemath{0.90}{\forall n\in \mathcal{N}, \forall m\in \mathcal{M}},\notag\\
&\mbox{C4:}~\mathbf{b}_n^T\mathbf{D}\mathbf{b}_{n'}\geq D_{\mathrm{min}},\ n\neq n',\ \forall n, n'\in\mathcal{N},\nonumber\\
    &\mbox{C5:}~\frac{1}{Q} \frac{1}{L} \sum_{q=1}^Q\sum_{l=1}^L \bigg|\rho_{0}\mathcal{D}(\theta_l,\phi_{q}) -\mathbf{\hat{a}}^H(\theta_l,\phi_{q}) \mathbf{B}\mathbf{R_{x}}\nonumber\\&\hspace{15mm}\mathbf{B}^{T}\mathbf{\hat{a}}(\theta_l,\phi_{q})\bigg|^2\leq \delta_{d},  \hfill\nonumber \\
  &\mbox{C6:}~\text{Pr}\bigg \{\Gamma_{e}\leq\Gamma^{\text{th}}_{e}\bigg \}\leq \nu,
\end{align}
    where $\text{C}{1}$ guarantees that the SINR of user $k$ does not fall bellow the minimum required threshold $\gamma^{\text{th}}_{k}$. $\text{C}{2}$ indicates that, for each MA element, only one position can be selected. $\text{C}{3}$ enforces that the position selection indicator for MA is an integer variable. $\text{C}{4}$ guarantees that the minimum distance between any pair of MA elements exceeds $D_{\min}$. $\text{C}{5}$ ensures that the MSE between the desired radar beam pattern and
the actual beam pattern of the transmitted signal does not
exceed a predefined threshold $\delta_{d}$. Finally, $\text{C}{6}$ ensures that the sensing SNR for target $e$ at the DFRC-BS falls below threshold, $\Gamma^{\text{th}}_{e}$, with a probability of not more than $\nu$. 
\vspace{-1mm}
	\subsection{Solution of the Optimization Problem}
Optimization problem $\mathcal{P}_{0}$ is non-convex due to the coupling between the optimization variables, binary constraint $\text{C3}$, binary quadratic constraint $\text{C4}$, and non-convex constraints $\text{C1}$ and $\text{C6}$. In general, it is very challenging to find a globally optimal solution for non-convex optimization problems. Here, we propose a low-complexity AO based iterative algorithm, which yields a suboptimal solution of $\mathcal{P}_{0}$. In particular, we first optimize the beamforming for the communication users and sensing targets, and then we optimize the position of the MA elements. 
\subsubsection{Beamforming Optimization for Sensing and Communication}
First, the positions of the MA elements are fixed, i.e., $\mathbf{B}=\mathbf{B}^{(t)}$, where $t$ denotes the iteration index of the AO algorithm, and the beamforming for both communication and sensing are optimized. To do so, we employ semidefinite programming (SDP) and define $\mathbf{W}_{k}=\mathbf{w}_{k}\mathbf{w}^{H}_{k}$ and $\widetilde{\mathbf{H}}_{k}=\hat{\mathbf{h}}^{H}_{k}\hat{\mathbf{h}}_{k}$. For notational simplicity, we further define $\mathbf{F}_{k}\triangleq\mathbf{B}\mathbf{W}_{k}\mathbf{B}^{T}$ and $\mathbf{Y}\triangleq\mathbf{B}\mathbf{R}\mathbf{B}^{T}$. As a result, the SINR constraint can be restated as follows:
\vspace{-3mm}
\begin{align}
  \overline{\mbox{C1}}:\frac{\text{Tr}(\widetilde{\mathbf{H}}_{k}\mathbf{F}_{k})}{\sum_{i\neq k}\text{Tr}(\widetilde{\mathbf{H}}_{k}\mathbf{F}_{i})+\text{Tr}(\widetilde{\mathbf{H}}_{k}\mathbf{Y})+\sigma^{2}_{k}}\geq \gamma^{\text{th}}_{k}. 
\end{align}
Considering C5, we initially rewrite the beam pattern as 
 $\mathcal{D}(\mathbf{p}_{m},\theta_{l},\phi_{q},\mathbf{F}_{k},\mathbf{Y})\triangleq\mathbf{\hat{a}}^H(\theta_l,\phi_{q}) \Big(\sum_{k\in \mathcal{K}}\mathbf{F}_{k}+\mathbf{Y}\Big)\mathbf{\hat{a}}(\theta_l,\phi_{q})$. Subsequently, the beam pattern MSE constraint is reformulated as follows:\\ \\
\begin{equation}
\overline{\text{C5}}:\frac{1}{Q} \frac{1}{L} \sum_{q=1}^Q\sum_{l=1}^L \bigg|\rho_{0}\mathcal{D}(\theta_l,\phi_{q}) -\mathcal{D}(\mathbf{p}_{m},\theta_{l},\phi_{q},\mathbf{F}_{k},\mathbf{Y})\bigg|^2\leq \delta_{d}.  
\end{equation}
Next, we tackle constraint $\text{C6}$. We first recast this constraint as follows:\\
\vspace{-2mm}
\begin{align}\label{27}
\text{C6}:\text{Pr}\bigg \{\Omega_{e}\leq \frac{{16\pi \Psi^4_{e}\sigma^{2}_{e}\Gamma^{\text{th}}_{e}}}{L^{2}_{0}\mathcal{D}(\mathbf{p}_{m},\theta_{e},\phi_{e},\mathbf{F}_{k},\mathbf{Y})}\bigg\}\leq \nu,
\end{align} 
To calculate the probability in \eqref{27}, we recall that $\Omega_{e}$ is exponentially distributed. Therefore, this probability can be calculated as 
$1-\text{exp}\big(-\frac{{16\pi \Psi^4_{e}\sigma^{2}_{e}\Gamma^{\text{th}}_{e}}}{L^{2}_{0}\mathcal{D}(\mathbf{p}_{m},\theta_{e},\phi_{e},\mathbf{F}_{k},\mathbf{Y})}\times \frac{1}{\Omega_{\text{av}}}\big)$.
Consequently, we can restate \eqref{27} as follows:
\begin{align}
\overline{\text{C6}}:\mathcal{D}(\mathbf{p}_{m},\theta_{e},\phi_{e},\mathbf{F}_{k},\mathbf{Y})\geq -\frac{{16\pi \Psi^4_{e}\sigma^{2}_{e}\Gamma^{\text{th}}_{e}}} {\ln (1-\nu)\Omega_{\text{av}}L_{0}^{2}}.
    \end{align}
  Finally, the resulting optimization problem for the communication and sensing beams can be rewritten as follows
\begin{align}
 \label{P3}
  &\hspace*{-2mm}\mathcal{P}_{1}:\underset{\mathbf{W}_{k},\mathbf{R},\rho_{0}}{\mino}\hspace*{2mm}\sum_{k\in\mathcal{K}}\text{Tr}(\mathbf{W}_k)\notag+\text{Tr}(\mathbf{R})\\
\mbox{s.t.}~~&\mbox{C7}:\text{Rank}(\mathbf{W}_{k})\leq 1,~\overline{\mbox{C1}},\overline{\mbox{C5}},\overline{\text{C}6}. 
\end{align}
Now, by removing rank-one constraint \text{C7} and employing SDP relaxation, problem $ \mathcal{P}_1 $ becomes a convex optimization problem, which can be effectively solved using CVX. The tightness of the SDP relaxation can be confirmed using a similar approach as in\cite[Appendix A]{Rank}. However, due to space limitations, the proof is omitted here.
\subsubsection{Optimization of Position of MA Elements}
In this subsection, we optimize the positions of the MA elements for given $\mathbf{W}_{k}=\mathbf{W}^{(t)}_{k}$ and $\mathbf{R}=\mathbf{R}^{(t)}$.  We start by reformulating the quadratic inequality constraint C4 into three linear inequality constraints using the following lemma\cite{glover1974converting}.
\begin{lemma} [See\protect{\cite{glover1974converting}}]
    Inequality constraint C4 can be reformulated as a set of linear inequality constraints using binary auxiliary variables $\phi_{n,n',i,j}$
\begin{align}
&\scalemath{0.9}{\mathrm{C4a:}\sum_{i\in\mathcal{M}}\sum_{j\in\mathcal{M}}D_{i,j}\phi_{n,n',i,j}\geq D_{\mathrm{min}},n\neq n',\forall n,n'\in\mathcal{N}},\\
&\scalemath{0.9}{\mathrm{C4b:}\phi_{n,n',i,j}\leq \min\left\{b_n[i],b_n[j]\right\},n\neq n', \forall n,n'\in\mathcal{N},\forall i,j\in\mathcal{M}},\\
&\scalemath{0.9}{\mathrm{C4c:}\phi_{n,n',i,j}\geq b_n[i]+b_n[j]-1,n\neq n', \forall n,n'\in\mathcal{N},\forall i,j\in\mathcal{M}}.
\vspace{-3mm}
\end{align} 
\end{lemma}
To simplify notation, we introduce binary vector $\boldsymbol{\phi}=[\phi_{1,2,1,1},\cdots,\phi_{n,n',i,j},\cdots,\phi_{N-1,N,M,M}]$, $n\neq n'$, $\forall n,n'\in\mathcal{N}$, and $\forall i,j\in\mathcal{M}$ collecting all binary auxiliary variables.
Next, we relax the integer variables into continuous ones and we rewrite them as $\text{C8a}: 0\leq b_{n}[m]\leq 1$, $\text{C8b}: \sum_{m=1}^{M}\sum_{n=1}^{N}b_{n}[m]-b^{2}_{n}[m]\leq 0$, and  $\text{C9a}: 0\leq \phi_{n,n',i,j}\leq 1$, $\text{C9b}: \sum_{i\in \mathcal{M}}\sum_{j\in \mathcal{M}}\sum_{n\in \mathcal{N}}\sum_{n'\in \mathcal{N}}\phi_{n,n',i,j}-\phi^{2}_{n,n',i,j}\leq 0$. Then, we linearized C8b and C9b via Taylor approximation\cite{Globecom2023}. To facilitate the solution design, we add $\text{C10}:\mathbf{F}_{k}=\mathbf{B}\mathbf{W}_{k}\mathbf{B}^{T}$ and $\text{C11}:\mathbf{Y}=\mathbf{B}\mathbf{R}\mathbf{B}^{T}$ as two additional constraints in the optimization problem. Next, in the following lemma, we transform equality constraints $\mbox{C10}$ and $\mbox{C11}$ into equivalent inequality constraints \cite[Appendix A]{6698281}.
\begin{lemma}[See\protect{\cite[Appendix A]{6698281}}]Equality constraints $\mbox{C10}$ and $\mbox{C11}$ become equivalent to the following inequality constraints by introducing auxiliary optimization variables $\mathbf{S}$, $\mathbf{T}$, $\mathbf{U}$, and $\mathbf{V}$ and applying Schur's complement:
\begin{eqnarray}
\mbox{C10a:}&\hspace*{1mm}\label{sdp}
   \begin{bmatrix}
        \mathbf{S} & \mathbf{F}_{k} & \mathbf{B}\mathbf{W}_{k}\\
        \mathbf{F}_{k}^H & \mathbf{T} & \mathbf{B}\\
    \mathbf{W}_{k}^H\mathbf{B}^T & \mathbf{B}^{T} & \mathbf{I}_{N}
    \end{bmatrix}&\succeq \mathbf{0},\\
\mbox{C10b:}&\hspace*{1mm}\label{DC}
    \mathrm{Tr}\left(\mathbf{S}-\mathbf{B}\mathbf{W}_{k}\mathbf{W}_{k}^H\mathbf{B}^{T}\right)&\leq0,\\
    \mbox{C11a:}&\hspace*{1mm}\label{sdp2}
    \begin{bmatrix}
        \mathbf{U} & \mathbf{Y} & \mathbf{B}\mathbf{R}\\
        \mathbf{Y}^H & \mathbf{V} & \mathbf{B}^T\\
    \mathbf{R}^H\mathbf{B}^T & \mathbf{B} & \mathbf{I}_{N}
    \end{bmatrix}&\succeq \mathbf{0},\\
   \mbox{C11b:}&\hspace*{1mm}\label{DC2}
    \mathrm{Tr}\left(\mathbf{U}-\mathbf{B}\mathbf{R}\mathbf{R}^H\mathbf{B}^{T}\right)&\leq0.
\end{eqnarray}
\end{lemma}We note that constraints $\mbox{C10a}$ and $\mbox{C11a}$ are  linear matrix inequality (LMI) constraints, whereas $\mbox{C10b}$ and $\mbox{C11b}$ present a challenge due to their difference of convex (DC) function structure. To address these new non-linearities, we employ Taylor approximation for the DC components in $\mbox{C10b}$ and $\mbox{C11b}$, transforming them into affine constraints as $\overline{\mbox{C10b}}:f_{1}(\mathbf{S})-g_{1,k}(\mathbf{B})\leq 0$ and $\overline{\mbox{C11b}}:f_{2}(\mathbf{U})-g_{2,k}(\mathbf{B})\leq 0$ where $f_{1}(\mathbf{S})$, $g_{1,k}(\mathbf{B})$, $f_{2}(\mathbf{U})$, and $g_{2,k}(\mathbf{B})$ are defined as follows
\begin{align}
&f_{1}(\mathbf{S})\triangleq\mathrm{Tr}(\mathbf{S}), f_{2}(\mathbf{U})\triangleq \mathrm{Tr}(\mathbf{U}), \\
 &g_{1,k}(\mathbf{B})\triangleq\mathrm{Tr}\left(\mathbf{B}^{(t)}\mathbf{W}_{k}\mathbf{W}_{k}^H\mathbf{B}^{(t)^T}\right)-\nonumber\\ &2\text{Re}\Bigg[\mathrm{Tr}\bigg((\mathbf{W}_{k}\mathbf{W}_{k}^H\mathbf{B}^{(t)^T})(\mathbf{B}-\mathbf{B}^{(t)})\bigg)\Bigg],\\
  &g_{2,k}(\mathbf{B})\triangleq\mathrm{Tr}\left(\mathbf{B}^{(t)}\mathbf{R}\mathbf{R}^H\mathbf{B}^{(t)^T}\right)-\nonumber\\&2\text{Re}\Bigg[\mathrm{Tr}\bigg((\mathbf{R}\mathbf{R}^H\mathbf{B}^{(t)^T})(\mathbf{B}-\mathbf{B}^{(t)})\bigg)\Bigg],
\end{align}
where $\mathbf{B}^{(t)}$ is the solution in the $t$-th iteration. Finally, we introduce penalty factors $\tau_{i}$, $\forall i\in\{1,2,3,4\}$, to incorporate $\overline{\mbox{C8b}}$ and $\overline{\mbox{C9b}}$, $\overline{\mbox{C10b}}$, and $\overline{\mbox{C11b}}$ into the objective function. In this case, the optimization problem at hand can be written as follows:
\vspace{-2mm}
\begin{align}
\label{Position_problem}
  &\hspace*{-4mm}\mathcal{P}_{2}:\underset{\mathbf{B},\mathbf{F}_{k},\mathbf{Y},\mathbf{S},\mathbf{T},\mathbf{U},\mathbf{V},\boldsymbol{\phi}}{\mino}\sum_{k\in\mathcal{K}}\text{Tr}(\mathbf{W}_k)\notag+\text{Tr}(\mathbf{R})+\nonumber\\&\tau_{1}\big(f_{1}(\mathbf{S})-\sum_{k\in\mathcal{K}}g_{1,k}(\mathbf{B})\big)+\tau_{2}\big(f_{2}(\mathbf{U})-\sum_{k\in\mathcal{K}}g_{2,k}(\mathbf{B})\big)+\nonumber\\&\tau_{3}\sum_{m=1}^{M}\sum_{n=1}^{N}\big(b_{n}[m]-b^{(t)}_{n}[m](2b_{n}[m]-b^{(t)}_{n}[m])\big)+\tau_{4}\nonumber\\&\hspace{-4mm}\sum_{i\in \mathcal{M}}\sum_{j\in \mathcal{M}}\sum_{n\in \mathcal{N}}\sum_{n'\in \mathcal{N}}\big(\phi_{n,n',i,j}-\phi_{n,n',i,j}^{(t)}(2\phi_{n,n',i,j}-\phi_{n,n',i,j}^{(t)})\big)\nonumber\\
\mbox{s.t.}~&\overline{\mbox{C1}},\mbox{C2},\mbox{C4a},\mbox{C4b},\mbox{C4c},\overline{\text{C5}},\overline{\text{C6}},\mbox{C8a},\mbox{C9a},\mbox{C10a},\mbox{C11a}.
\end{align}
In each iteration $t$, we update the solution set and efficiently solve $\mathcal{P}_{2}$ by CVX. The proposed solution based on AO is summarized in \textbf{Algorithm} \ref{algorithmo}. Note that for sufficiently large penalty factors, $\tau_{i},\forall i\in \{1,2,3,4\}$, in $\mathcal{P}_{2}$, the objective function of $\mathcal{P}_{0}$ is non-increasing in each iteration of \textbf{Algorithm 1} and converges to a high-quality suboptimal solution with polynomial time computational complexity \cite{Globecom2023}. The computational complexity of \textbf{Algorithm 1} can be shown to be given by $\mathcal{O}\Big(\mathrm{log}(1/\varepsilon_{\text{AO}})\big((2K+E+1)N^{3}+(2K+E+1)^{2}N^2+(K+2N+10MN+KMN+E+1)N^{3}+(K+2N+10MN+KMN+E+1)N^{2}\Big)$, where $\mathcal{O}\left ( \cdot  \right )$ is the big-O notation and $\varepsilon_{\text{AO}}$ is the convergence tolerance of \textbf{Algorithm 1}. 
\begin{algorithm}[t]
    \footnotesize
    \captionof{algorithm}{Proposed resource allocation framework.}
     \label{algorithmo}
     1.\quad Initialize  $\mathcal{F}^{(0)}$, $\mathbf{B}^{(0)}$, $\tau_{i}\gg 1$, $\forall i\in\{1,2,3,4\}$,~$t=0$,~$\varepsilon_{\text{AO}}$.\\
     \textbf{Repeat} \\
      2.\quad Solve $\mathcal{P}_1$ for given $\mathbf{B}=\mathbf{B}^{(t)}$ and obtain $\mathbf{w}_{k}^{(t)}$ and $\mathbf{R}^{(t)}$.\\
3. \quad Solve $\mathcal{P}_2$ for given $\mathbf{w}_{k}= \mathbf{w}_{k}^{(t+1)}$, $\mathbf{R}= \mathbf{R}^{(t+1)}$, and obtain $\mathbf{B}^{(t+1)}$.\\
4. \quad Set~$t=t+1$\\
       5. \textbf{until}~$\frac{\mathcal{F}^{(t)}-\mathcal{F}^{(t-1)}}{\mathcal{F}^{(t-1)}}\leq \varepsilon_{\text{AO}}$. 
\end{algorithm}

\section{Simulation Results}
\vspace{-1mm}
In this section, we evaluate the performance of MA-enabled ISAC systems and study the impact of RCS fluctuations via numerical simulations. The transmitter area is modeled as a square with dimensions $a\lambda \times a\lambda$, where $a$ represents the normalized size of the transmitter area at the BS. The default parameters are detailed in Table~\ref{tab:simulation_parameters}. The users are randomly distributed, with the distance from the BS uniformly ranging from $10$ m to $50$ m, while two sensing targets are positioned at $(\theta_{1}=0^\circ,\phi_{1}=0^\circ)$ and $(\theta_{2}=30^\circ,\phi_{2}=30^\circ)$, with their distances from the BS varying between $10$ m and $25$ m.  
We compare our proposed approach against two baseline schemes. In baseline scheme 1, the positions of the antenna elements are fixed. The beamforming vectors are determined by solving the beamforming problem for a fixed antenna array. 
Baseline scheme 2 utilizes an AS strategy, where the BS is equipped with a $2\times N$ uniform planar array (UPA) with fixed-position antennas, spaced $\lambda/2$ apart. This setup ensures statistically independent channels across different antennas. The beamforming optimization is conducted for all possible subsets of $N$ antenna elements, and the subset that achieves the lowest BS transmit power is selected.
\begin{table}[t]
\caption{Simulation Parameters}
\vspace{-2mm}
\label{tab:simulation_parameters}
\centering
\setlength{\tabcolsep}{6pt}
\begin{tabular}{|c|c|}
\hline
\textbf{Parameter} & \textbf{Value} \\
\hline
$N$ & 4 \\
\hline
$K$ & $2$ \\
\hline
$E$ & $2$ \\
\hline
Carrier frequency & 5 GHz \\
\hline
Wavelength ($\lambda$) & 0.06 m \\
\hline
Normalized transmitter size ($a$) & $2$ \\
\hline
Path-loss exponent & $\alpha=2.2$\\
\hline
Large scale fading ($L_{0}$)  & $-30$~dB\\
\hline
$D_{\text{min}}$ & 0.015 m \\
\hline
$\sigma^{2}_{k}=\sigma^{2}_{e}$ & -80 dBm \\
\hline
$d$ & $0.01$ m \\
\hline
$\nu$ & $0.01$\\
\hline
$\gamma^{\text{th}}_{k}$ & $10$ dB\\
\hline
$\Gamma^{\text{th}}_{e}$ & $10$ dB\\
\hline
$\Omega_{\text{av}}$ & $1$ m$^{2}$\\
\hline
\end{tabular}
\end{table}

Fig. 2 shows that the average transmit power of the DFRC-BS monotonically increases with the sensing SNR requirement for all considered schemes, reflecting the higher power needed to meet stricter sensing requirements. Moreover, Fig. 2 reveals the effectiveness of MA positioning in reducing the required transmit power, especially when compared to the fixed antenna configuration of baseline scheme 1, which leads to suboptimal spatial correlations of the transmit signal. In contrast, the proposed approach selects the optimal MA positions for the given channel conditions. Although baseline scheme 2, through AS, improves the spatial DoFs at the DFRC-BS, its limited choices for the antenna positions, spaced at $\lambda/2$, still restrict the spatial resolution. These findings highlight that MAs with their sub-wavelength precision can significantly enhance the system's ability to shape its radiation pattern benefiting both communications and sensing, thereby boosting the power efficiency of ISAC systems. Furthermore, by varying the maximum tolerable probability of failure, $\nu$, we observe that meeting stricter sensing QoS requirements, e.g., $\nu=0.01$, requires a higher transmit power compared to more relaxed sensing QoS requirements, e.g., $\nu=0.1$. Notably, the proposed scheme with optimized MA positions consumes less power for the stricter sensing QoS requirement of $\nu=0.01$ than baseline scheme 1 with fixed antenna positioning for the more relaxed QoS sensing requirement of $\nu=0.1$. This not only demonstrates the efficiency of the proposed scheme but also highlights its ability to mitigate the challenges caused by the dynamic RCS. 


Fig. 3 depicts the average transmit power versus the normalized transmitter size. For the proposed scheme, as the normalized transmitter size increases, the average transmit power required decreases due to the increased flexibility in MA positioning. Furthermore, for relaxed sensing QoS requirements (i.e., $\nu=0.1$), the proposed scheme continues to benefit from increasing the transmitter size also for $a>3$, whereas performance saturates for $a>3$ for stricter sensing QoS requirements (i.e., $\nu=0.01$). In other words, the range of transmitter areas in which the proposed scheme can benefit from the additional DoFs offered by flexible MA positioning is larger for less stringent sensing QoS requirements. On the other hand, baseline schemes 1 and 2 cannot benefit from an expanded transmitter area, as their performance is limited by their fixed antenna spacing of half a wavelength.
 \vspace{-2mm}
\section{Conclusion}
\vspace{-1mm}
In this paper, we investigated for the first time MA-assisted multi-user multi-target ISAC systems with spatially discrete antenna positions. We introduced a new performance metric for evaluating the sensing QoS, employing a chance constraint to manage the impact of the uncertainty introduced by dynamic RCS variations. Subsequently, we optimized the MA positions and the communication and sensing beams for minimization of the total transmit power at the BS while ensuring the individual communication and sensing
task QoS requirements. The resulting non-convex problem was solved via AO where a high-quality suboptimal solution was obtained. Simulation results revealed that the proposed approach enables substantial power savings in ISAC systems, primarily due to the optimization of the MA positions, compared to baseline schemes employing fixed antenna positions and AS. Our results further unveiled that the sub-wavelength positioning of MA elements enhances the robustness of ISAC systems against RCS fluctuations, particularly under stringent sensing QoS constraints.
\begin{figure}[t]
    \centering
\includegraphics[width=5.500cm]{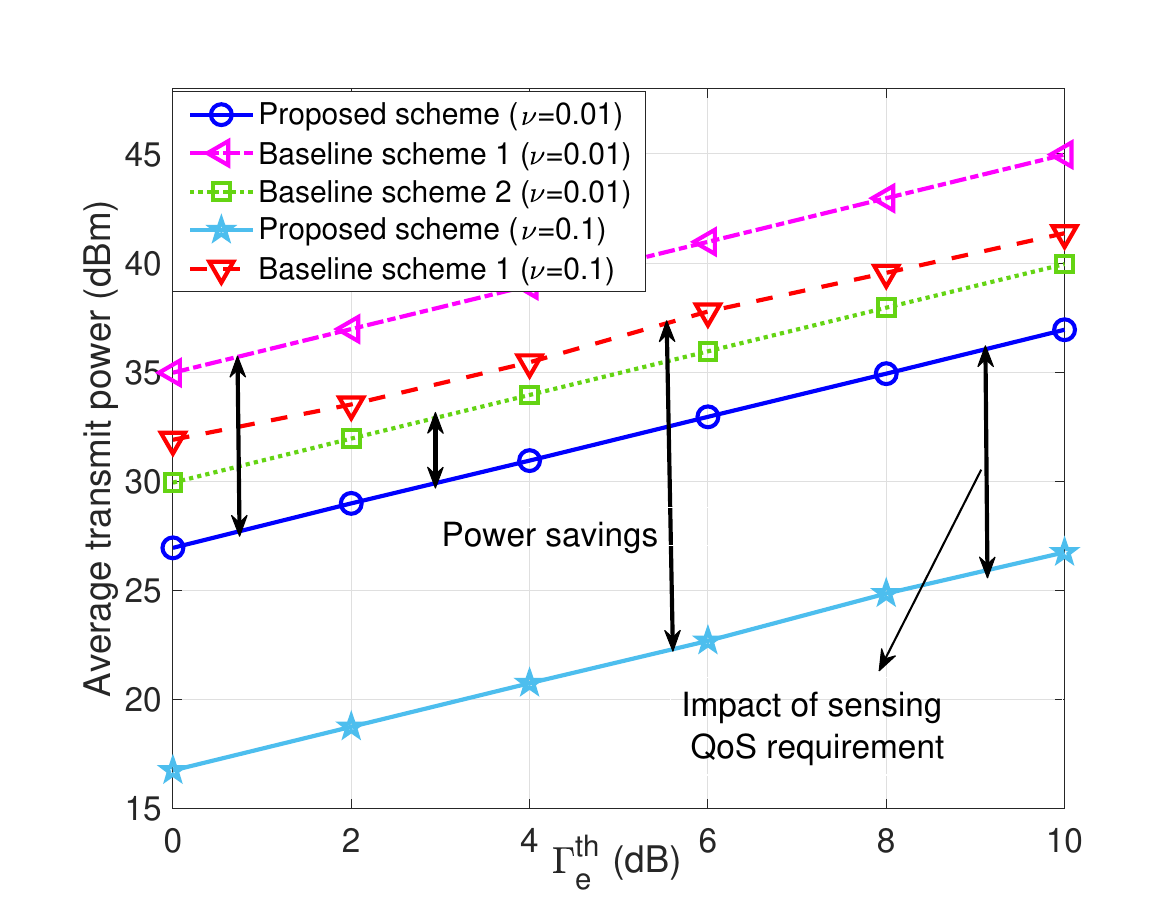}
\vspace{-2mm}
\caption{Average transmit power versus minimum required sensing SNR.}
\vspace{-5mm}
\end{figure}
\begin{figure}[t]
    \centering
\includegraphics[width=5.500cm]{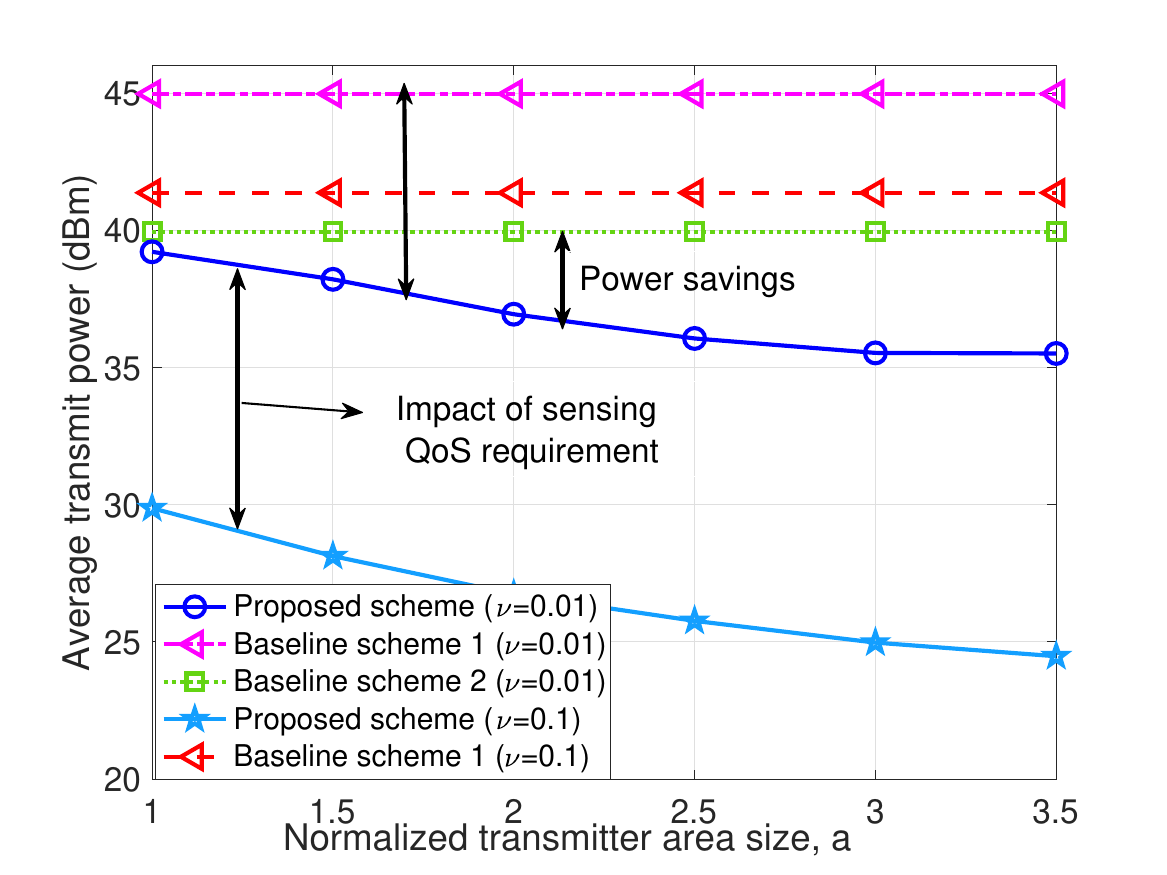}
\vspace{-3mm}
\caption{Average transmit power versus normalized transmitter area size.}
\end{figure}
\vspace{-3mm}
\bibliographystyle{IEEEtran}
\bibliography{reference.bib}
\end{document}